\documentclass[a4paper,12pt]{article}

\usepackage{tabularx, booktabs, ragged2e}
\newcolumntype{L}{>{\RaggedRight\arraybackslash}X}

\usepackage{graphicx}
\usepackage[english]{babel} 
\usepackage[utf8]{inputenc}
\usepackage{indentfirst}
\usepackage{textcomp}
\usepackage{float}
\usepackage[labelfont=bf, textfont=small,
			justification=justified,
			singlelinecheck=false, margin=2cm]{caption}
\usepackage[labelfont=bf, textfont=small,  justification=justified, singlelinecheck=false, margin=2cm]{subcaption}
\usepackage[caption2]{ccaption} 
\usepackage{multirow}
\usepackage{amsmath}
\usepackage[sfdefault]{roboto}

\usepackage{multicol}
\usepackage[table]{xcolor} 
\usepackage{hyperref} 
\hypersetup{colorlinks, citecolor=black,
   		 	filecolor=black, linkcolor=black,
    		urlcolor=black}
\usepackage{authblk} 

\usepackage{natbib} 
\usepackage{nameref}

\usepackage[top=2cm, bottom=2cm, inner=1.5cm, outer=1.5cm]{geometry}

\begin{document}

\title{GloBIAS: strengthening the foundations of BioImage Analysis}

\author[a,$\dagger$]{Agustin A. Corbat}
\author[b,c,$\dagger$]{Christa  G.  Walther}
\author[d,e,$\dagger$]{Laura R. de la Ballina}
\author[f,$\ddagger$]{Nicholas D. Condon}
\author[g,$\ddagger$]{Alessandro A. Felder}
\author[h,i,$\ddagger$]{Martin Schätz}
\author[j,$\ddagger$]{Bettina Schmerl}
\author[k,$\ddagger$]{Ko Sugawara}
\author[l]{Clara Prats}
\author[a]{Anna Klemm}
\author[m,n]{Florian Levet}
\author[o]{Kota Miura}
\author[p]{Paula Sampaio}
\author[q]{Christian Tischer}
\author[r,s,*]{Rocco D'Antuono}
\author[t,*]{Beth A Cimini}
\author[u,*]{Robert Haase}

\affil[a]{BioImage Informatics Unit, Science for Life Laboratory and Department of Information Technology, Uppsala University, Sweden}
\affil[b]{German BioImaging – Gesellschaft für Mikroskopie und Bildanalyse e.V., Konstanz, Germany}
\affil[c]{University of Vienna, Vienna, Austria}
\affil[d]{Centre for Cancer Cell Reprogramming, Institute of Clinical Medicine, Faculty of Medicine, University of Oslo, Montebello, 0379, Oslo, Norway}
\affil[e]{Department of Molecular Cell Biology, Institute for Cancer Research, The Norwegian Radium Hospital, Oslo University Hospital, Montebello, 0379, Oslo, Norway}
\affil[f]{Institute for Molecular Bioscience, The University of Queensland, Brisbane, Queensland, Australia}
\affil[g]{Sainsbury Wellcome Centre \& Advanced Research Computing, University College London, London, United Kingdom}
\affil[h]{Vinicna Microscopy Core Facility, Faculty of Science, Charles University, Prague, Czech Republic}
\affil[i]{Department of Mathematics Informatics and Cybernetics, University of Chemistry and Technology, Prague, Czech Republic}
\affil[j]{Picower Institute for Learning and Memory, Massachusetts Institute of Technology, Cambridge MA, USA}
\affil[k]{Laboratory for Developmental Dynamics, RIKEN Center for Biosystems Dynamics Research, Kobe, Japan}
\affil[l]{University of Copenhagen}
\affil[m]{Univ. Bordeaux, CNRS, Interdisciplinary Institute for Neuroscience, IINS, UMR 5297, F-33000 Bordeaux, France}
\affil[n]{Univ. Bordeaux, CNRS, INSERM, Bordeaux Imaging Center, BIC, UAR3420, US 4, F-33000 Bordeaux, France}
\affil[o]{Bioimage Analysis \& Research, Heinrich-Fuchsstr. 29, Heidelberg, 69126 Heidelberg Germany}
\affil[p]{i3S - Instituto de Investigação e Inovação em Saúde, Universidade do Porto, Porto, Portugal}
\affil[q]{EMBL, Data Science Centre, Heidelberg, Germany}
\affil[r]{Crick Advanced Light Microscopy STP, The Francis Crick Institute, London, United Kingdom NW1 1AT}
\affil[s]{Department of Biomedical Engineering, School of Biological Sciences, University of Reading, Reading, UK - RG6 6AY}
\affil[t]{Imaging Platform, Broad Institute of MIT and Harvard, Cambridge MA, USA}
\affil[u]{Center for Scalable Data Analytics and Artificial Intelligence (ScaDS.AI) Dresden/Leipzig, Universität Leipzig, Leipzig, Germany }

\affil[$\dagger$]{Authors contributed equally}
\affil[$\ddagger$]{Authors contributed equally and were ordered alphabetically}
\affil[*]{Corresponding authors: rocco.dantuono@crick.ac.uk, bcimini@broadinstitute.org, robert.haase@uni-leipzig.de}

\date{}
\maketitle

\abstract{
There is a global need for BioImage Analysis (BIA) as advances in life sciences increasingly rely on cutting-edge imaging systems that have dramatically expanded the complexity and dimensionality of biological images. Turning these data into scientific discoveries requires people with effective data management skills and knowledge of state-of-the-art image processing and data analysis, in other words, BioImage Analysts. The Global BioImage Analysts’ Society (GloBIAS) aims to enhance the profile of BioImage Analysts as a key role in science and research. Its vision encompasses fostering a global network, democratising access to BIA by providing educational resources tailored to various proficiency levels and disciplines, while also establishing guidelines for BIA courses. By collaboratively shaping the education of BioImage Analysts, GloBIAS aims to unlock the full potential of BIA in advancing life science research and to consolidate BIA as a career path. To better understand the needs and geographical representation of the BIA community, a worldwide survey was conducted and 291 responses were collected across people from all career stages and continents. This work discusses how GloBIAS aims to address community-identified shortcomings in work environment, funding, and scientific activities. The survey underscores a strong interest from the BIA community in activities proposed by GloBIAS and their interest to actively contribute. With 72\% of respondents willing to pay for membership, the community's enthusiasm for both online and in-person events is set to drive the growth and sustainability of GloBIAS.
}

\vspace{2cm}


\newpage

\setlength\parskip{.5cm} 

\section*{Introduction}

For the past twenty years, the speed of innovation in bioengineering, labelling and imaging has tremendously accelerated, with today’s experiments producing exponentially increasing amounts of data to be analyzed (\cite{Williams2017}, \cite{Hartley2022}; \cite{Ruan2024}; \cite{Kyoda2024}). Bioimage Analysis has become a crucial step between image acquisition and scientific insight that requires specific data analysis knowledge and experience to ensure the correct interpretation of biological phenomena (\cite{Cimini2024}). As BioImage Analysts originate from 3 main disciplines - biology, physics and computer science – each with its own vocabulary and expectations, the field was in a dire need of structuration, communication and harmonization.

The Network of European BioImage Analysts (NEUBIAS) was established in 2016 (supported through EU funded \href{https://www.google.com/url?q=https://www.cost.eu/actions/CA15124/%23tabs%2BName:Description&sa=D&source=docs&ust=1747835717065848&usg=AOvVaw2iQ9gzgK34qmiFxSwwrvCF}{COST Action 15124}) and constituted the first community effort undertaken to shape the BioImage Analysis field and successfully established the term BioImage Analyst (\cite{Martins2021}, \cite{Miura2016}). NEUBIAS also welcomed International Partner Countries (Australia, Singapore, USA and Brazil) fostering intense discussions on the future of BioImage Analysis and Analysts at the global level. Importantly, the NEUBIAS philosophy disseminated outside of Europe and impacted several bioimaging communities such as BioImaging North America (BINA), Latin America BioImaging (LABI), Singascope, and Royal Microscopy Society (RMS) (\cite{DeNiz2024}). The Global BioImage Analysts’ Society (GloBIAS) stems from the success of the NEUBIAS activities and thus aims to build upon the same spirit and discipline to create a global and sustainable effort gathering the entire BioImage Analysis community, irrespective of their country or continent of origin. To fulfill this need, GloBIAS, a non-profit association now established as a formal society, aims to enable the sustainable  operation of the community and to connect  scientists involved in BioImage Analysis worldwide.

In order to make sure that GloBIAS covers the needs of BioImage Analysts across the globe, a community survey was conducted in 2024 to gain insight into several aspects of the BioImage Analyst position worldwide. In this paper, we report the current situation of the profession based on the results of this survey, map the needs and interests of the global BioImage Analysis community, and present the mission statement of GloBIAS.

\newpage
\section*{Description of the survey}

The goal of the survey was to gain insight into the community of people working in BioImage Analysis and seeing themselves as BioImage Analysts. The survey consisted of a collection of 6 demographic questions and 22 questions addressing topics such as funding, work description, activities that would be of interest, willingness to financially or actively contribute, and more (see \nameref{section:supplementary} and \href{https://github.com/GloBIAS-BioimageAnalysts/Survey_2024/blob/main/Survey24%20question%20list(1).docx}{repository link}). The survey was open between February 7th and April 29th 2024, shared as a Google form and distributed via social media (X, LinkedIn), imaging community newsletters (e.g. NEUBIAS, BINA, RMS, BioImaging UK, Light Microscopy Australia, LABI, ABIC), personal emails and networks (especially from Steering committee members), and list servers (e.g. Confocal list server, Image.sc).

\section*{Demographics of respondents and their current state}

\begin{figure}[!htbp]
    \centering
    \includegraphics[width=6.5in]{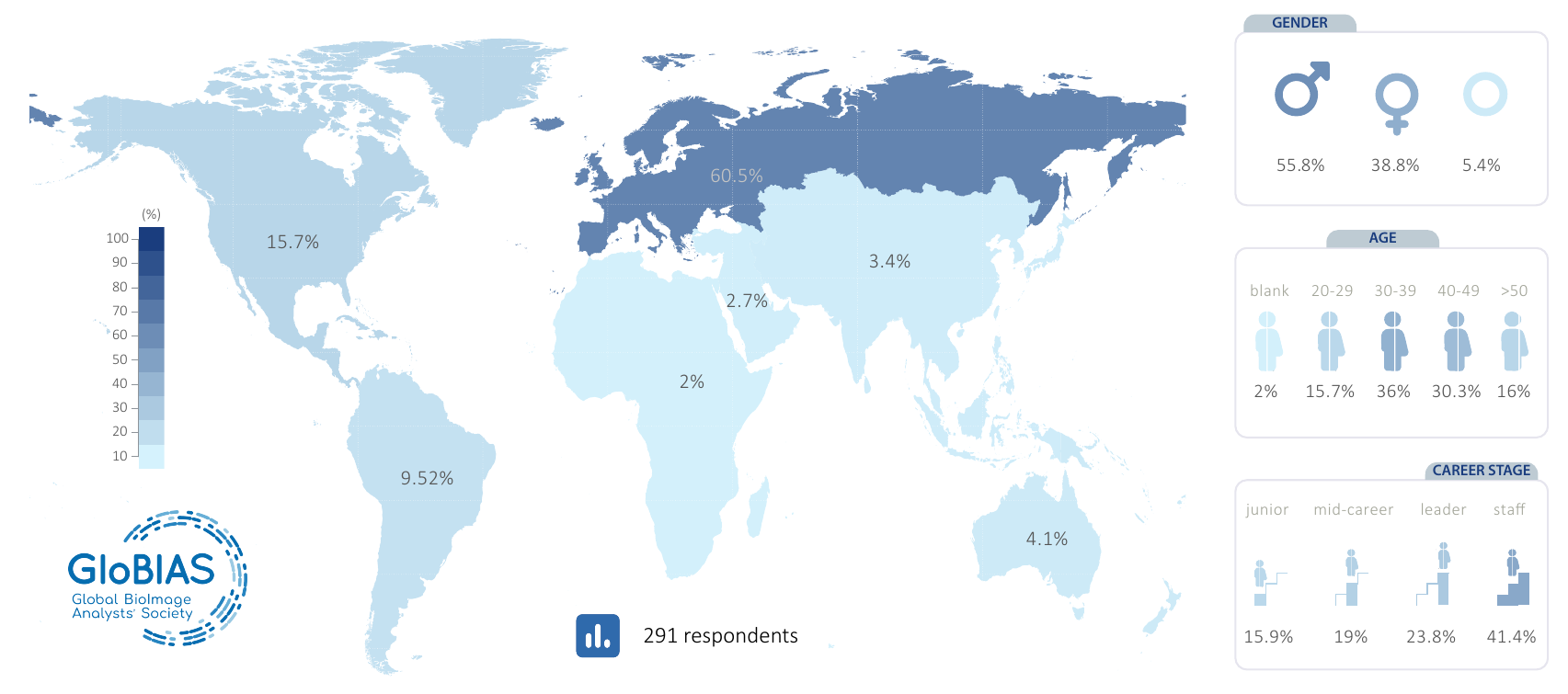}
    \caption{\textbf{Demographics of the survey responses}. Map representing the percentages of the 291 responses coming from people working in each geographical region. Specifically, there were 178 (60.5\%) from Europe, 47 (15.7\%) from North America, 28 (9.52\%) from South America, 12 (4.1\%) from Australia/Oceania, 18 (3.4\%) from Asia and 6 (2\%) from Africa. With regards to gender, there were 164 (55.8\%) male respondents, 114 (38.8\%) female respondents, and 10 (5.4\%) respondents who preferred not to say or left blank. With regards to age groups, there were 46  (15.7\%) of respondents between 20 and 29, 106 (36\%) between 30 and 39, 89 (30.3\%) between 40 and 49, and 47 (16\%) above 50 years old. After classifying the positions into career stages, 46 (15.9\%) of respondents had a junior role, 55 (19\%) had a mid-career level, 69 (23.8\%) had a leader role and 120 (41.5\%) had a staff position. }
    \label{figure:demographics}
\end{figure}

The distribution of demographics showed that while most of the 291 respondents were based in Europe (61\%), all geographical regions were represented (see Figure \ref{figure:demographics}\textbf{A}). With regards to gender, 55.8\% identified as male, 38.8\% as female, with a small number electing not to say, left blank, or selecting "other" (5.4\%). The role of the BioImage Analysts is varied, with instruction, service, leadership of own projects, management, tool development, and infrastructure, each typically taking between 10-25\% of people’s time. About 60\% of respondents had a permanent position (see Supplementary Figure \ref{supplementary_figure:demographics}). Together, these data suggest a successful, possibly typically early-to-mid career core of BioImage Analysts mostly in Europe and the Americas, connected with motivated individuals located elsewhere. The latter may be particularly well-placed to further the society’s global reach.

\section*{Interests from the community}

\begin{figure}[!htbp]
    \centering
    \includegraphics[width=6.5in]{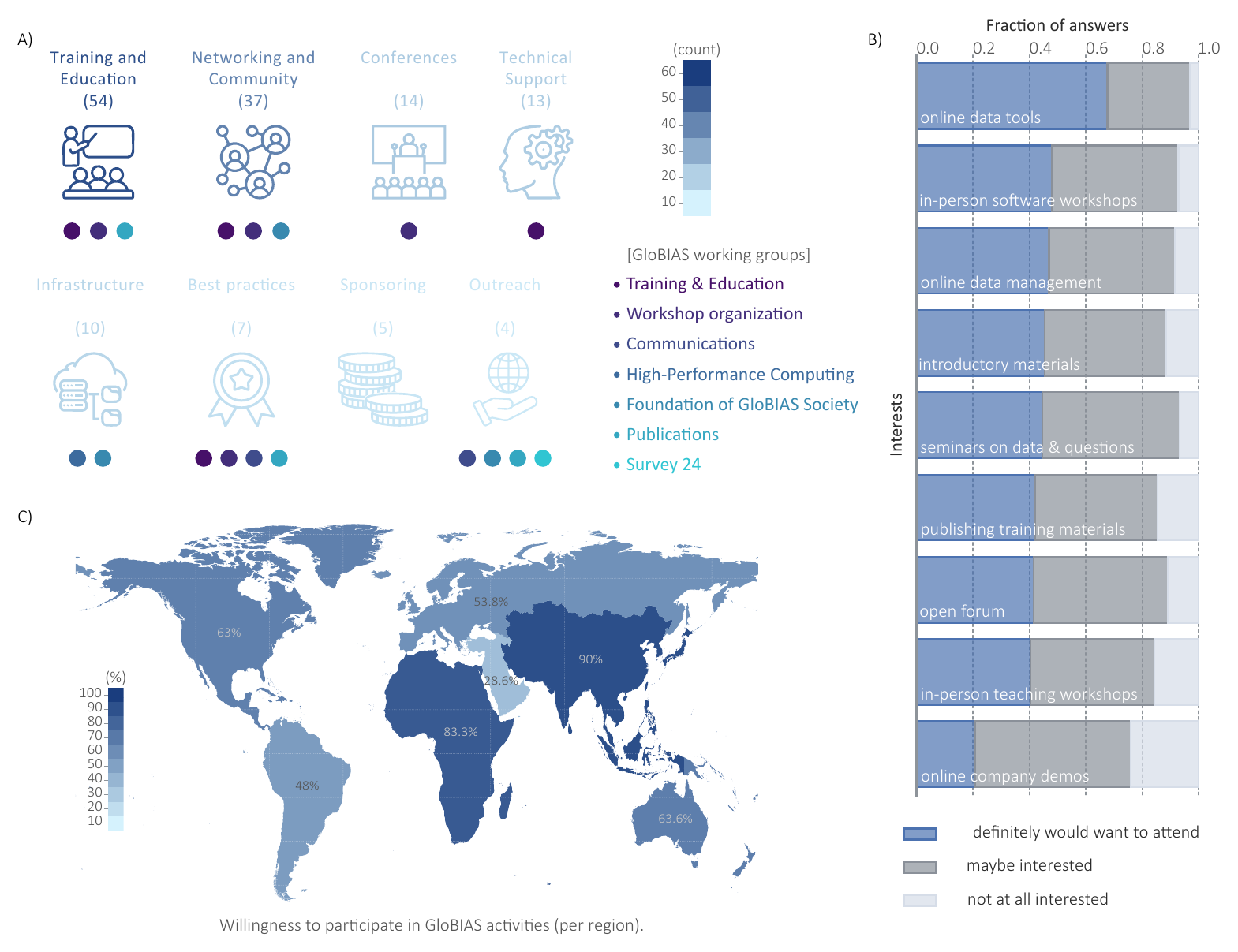}
    \caption{\textbf{GloBIAS working groups’ activities as a response to the community interests. A)} count of survey free text responses expressing interest in activities related to Training and Education (54), Networking and Community (37), Conferences (14), Technical Support (13), Infrastructure (10), Best practices (7), Sponsoring (5), and Outreach (4). Below each topic, GloBIAS working groups that are already involved in activities related to the aforementioned interests. \textbf{B)} Stacked bar plot showing the fraction of respondents indicating their interest level in different proposed activities. \textbf{C)} World map depicting the interest to actively contribute in GloBIAS activities, comparing responses from the different regions. Positive responses for each region as follows: Europe, 93 out of 173 (53.76\%); North America, 28 out of 45 (62.22\%);  South America, 12 out of 25 (48.00\%); Australia/Oceania, 7 out of 11 (63.64\%); Asia, 9 out of 10 (90.00\%); Africa, 5 out of 6 (83.33\%) and Near/Middle East, 2 out of 7 (28.57\%)}
    \label{figure:interests}
\end{figure}

To gather information on the various needs across the global community, we first collected open-ended feedback on how GloBIAS could better support the global community, and then explored interest in specific formats and topics. The responses demonstrate a strong interest in training and skill development, with a desire for learning opportunities ranging from introductory to advanced levels, and covering specific areas such as statistics, machine learning and deep learning. Networking and community building are also highly valued, highlighting the importance of connection, collaboration, knowledge exchange and communication. Conferences are recognised as valuable opportunities for training, networking and knowledge exchange. Technical support is a significant concern, with the community seeking guidance and assistance from experts, particularly for project-specific questions and data management. Furthermore, the community requires robust infrastructure to support their activities, and is committed to best practices, standards and open-source methodologies. Finally, outreach and sponsoring are considered important to promote accessibility and facilitate participation in events and training programmes (see Figure \ref{figure:interests}\textbf{A}). Some unique responses worth highlighting underscore the distinct regional needs of collaborative research (Africa), tool development (Australia/Oceania), hackathons (Europe), career development (Europe and North America), data management (Australia/Oceania and North America), quality control (South America), and financial support (South America). There was only one response from Asia, asking for "Basic Image Analysis tasks and algorithms using Whole Slides". 

When asked which would be the most interesting GloBIAS activities for the next year, 97\% of respondents showed interest in online events about new BioImage Analysis tools, being the most popular. There was also significant interest (90\%) in in-person workshops on new tools, and online sessions about infrastructure and managing image data, as well as exploring open questions and datasets in the field (see Figure \ref{figure:interests}\textbf{B}). Regarding the frequency of online events, most people prefer them to happen every three months.

Amongst the responses, several unique and interesting suggestions emerged for potential events in the field. Notably, there is a call for teaching good practices and offering introductory BIA courses, which could be highly beneficial for the BIA community. Additionally, quality control and fraud prevention were highlighted as important areas to address. In-person events focusing on image data infrastructure and management, as well as the formation of a core facilities group, were also mentioned. Respondents also expressed interest in organizing meet and greet retreats for researchers to foster connections. Furthermore, the idea of creating a BioImage Analysis journal was proposed.

On the other hand, some areas were identified as already established but in need of more attention and promotion. These include curating a tools catalogue (\href{http://biii.eu}{biii.eu}), covering the fundamentals of bioimaging (\href{https://www.bioimagingguide.org/}{Bioimaging Guide}), and creating easy-to-use tools to develop and deploy workflows (such as Galaxy, NextFlow, MCMICRO, or others).

The voiced interests reflect the full span of skills and knowledge needed to tackle the future of BIA, with the different aspects only to be covered by involving a range of active members.

\section*{Respondents show interest to actively contribute to GloBIAS}

In order to achieve the goals and mission of the GloBIAS initiative and support the global BioImage Analyst community, active involvement of its members is essential and 56.5\% indicated their interest in helping to organize and run regular events. Additionally, when asked in which working group of GloBIAS survey participants would like to join, 65\% of the respondents selected at least one, often multiple. More than 50 volunteers at the time of writing the manuscript joined one or two working groups (see \href{https://www.globias.org/activities/working-groups}{GloBIAS working groups webpage}). Reflecting the community's priorities, the \textit{Training \& Education} and \textit{Workshop 2024} working groups have garnered the most significant initial membership, with over 10 members each. Survey participants indicated equal interest in contributing to working groups hosting and creating material for training events as well as establishing criteria for scientific data integrity in BioImage Analysis (104 responses each of 286 respondents). Slightly fewer indicated interest in working groups to prepare online events and annual meetings (90 responses) or in-person events and hackathons (80 responses). Only 57 respondents expressed willingness to engage in outreach activities of the society such as social media channels and conference presentations. When queried about factors that would encourage further contribution, the majority of BioImage Analysts cited training opportunities as key. Open communication channels, online events, and alignment with scientific expertise were also frequently mentioned, with an open help desk noted by many. Respondents from all over the globe expressed their interest to actively contribute in society activities as shown in Figure \ref{figure:interests}\textbf{C}. Although the number of responses in some regions is concerning, this probably does not reflect the overall interest in BioImage Analysis in this region but the role of BioImage Analyst being barely established.

\section*{GloBIAS’s vision and goals as a response to the community needs}

In response to the community-identified needs, GloBIAS has emerged as a global hub for BioImage Analysis, uniting a diverse community of analysts, developers, scientists, and educators. The broad spectrum of backgrounds brings valuable skills and experiences into the community from all directions in science and engineering. From BioImage Analysts, bioinformaticians and research software engineers to life scientists, biologists and physicists worldwide, our community collaborates to advance the field.

\begin{figure}[!htbp]
    \centering
    \includegraphics[width=6.5in]{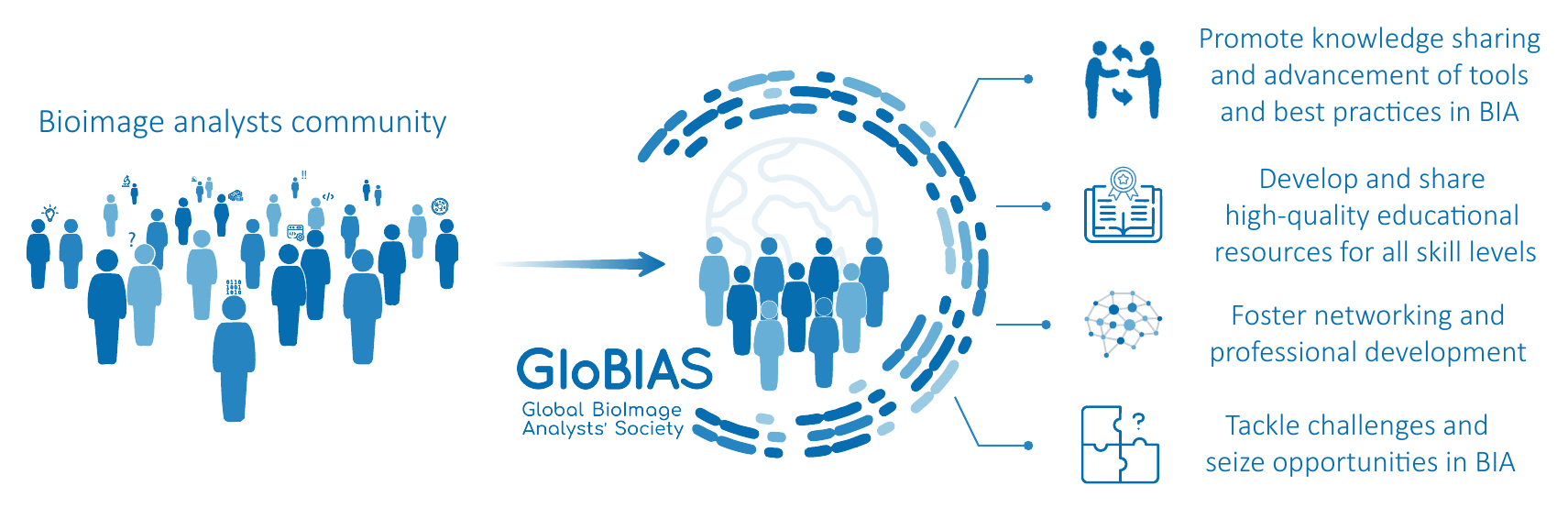}
    \caption{\textbf{GloBIAS vision and goals emerge from the BioImage Analysts community as a response to their needs.}}
    \label{figure:goals}
\end{figure}

The core mission of GloBIAS is to enhance our capabilities in "Quantitatively measuring biological systems using image data", encompassing the following specific goals: 1) fostering knowledge exchange, collaboration, advancement of tools and the development of best practices and standards in BioImage Analysis; 2) creating and disseminating accessible, high-quality educational resources for all skill levels; 3) facilitating networking and professional growth through online and in-person events; 4) proactively addressing emerging challenges and opportunities in BioImage Analysis (see Figure \ref{figure:goals}). Despite BioImage Analysis being a specialized discipline, there is currently no academic degree nor defined career path. Most analysts were originally trained in other fields (e.g. computer sciences, physics or biology) and independently acquired their skills. While this interdisciplinarity enriches the field, it also presents obstacles to career advancement and recognition—issues that GloBIAS is committed to resolving.

While only officially registered as an association in October 2024, GloBIAS has been actively engaged in realizing its mission since its funding by CZI (grant no 2023-321238, hosted by German BioImaging) in January 2023. After organizing the different working groups, several activities emerged and are currently ongoing, such as: 1) in-person events (GloBIAS in-person workshops 2024 in Sweden and 2025 in Japan), 2) creation and maintenance of databases of training materials, BioImage Analysts and trainers, 3) developing model training schools, including a Data Carpentries workshop, 4) Periodic online events (GloBIAS Seminar Series and consultancy drop-in sessions), 5) consolidating standards for BioImage Analysis, 6) implementing communication tools to ensure the successful engagement of the community, and 7) building and establishing of GloBIAS itself as a legal society and enabling the sustainable operation of the BioImage Analysis community.

Moving forward, GloBIAS is looking into broadening its reach and fostering inclusivity, particularly within regions currently underrepresented in the survey. To support this crucial role and ensure continuation of its activities beyond the initial round of funding, we have introduced a membership model with regional prices. By welcoming a diverse global membership, GloBIAS seeks to democratize access to BioImage Analysis resources and create a platform for the cross-pollination of ideas and solutions to regional challenges. We encourage you to connect with us and participate in this vibrant, international association. Visit the GloBIAS’s webpage \href{https://www.globias.org/home}{https://www.globias.org/home}) to join our community or subscribe to our newsletter.

\newpage
\section*{Acknowledgements}

The authors would like to thank all the GloBIAS volunteers, members, and contributors to the society, as well as all survey respondents, without whom this work would not be possible.

All figures have been designed using modified resources from \href{https://www.flaticon.com/}{Flaticon.com}.

\section*{Author Contributions}


\textbf{CRediT taxonomy}: \textbf{Conceptualization}: CG Walther, C Prats, F Levet, A Klemm, R D'Antuono, BA Cimini, R Haase; \textbf{Data curation}: CG Walther, AA Felder, BA Cimini, R Haase; \textbf{Analysis}: AA Corbat, CG Walther, ND Condon, AA Felder, M Schätz, B Schmerl, BA Cimini; \textbf{Funding acquisition}: C Prats, F Levet, P Sampaio, C Tischer, R D'Antuono, BA Cimini, R Haase ; \textbf{Methodology}: AA Felder, F Levet, BA Cimini, R Haase; \textbf{Project administration}: CG Walther, C Prats, C Tischer; \textbf{Software}: AA Corbat, AA Felder, M Schätz, BA Cimini; \textbf{Supervision}: AA Corbat, CG Walther, C Tischer, R D'Antuono, R Haase; \textbf{Validation}: AA Corbat, K Sugawara; \textbf{Visualization}: AA Corbat, LR de la Ballina, ND Condon; \textbf{Writing – original draft}: AA Corbat, CG Walther, LR de la Ballina, ND Condon, AA Felder, M Schätz, B Schmerl, F Levet, R D'Antuono, R Haase; \textbf{Writing – review \& editing}: AA Corbat, CG Walther, LR de la Ballina, ND Condon, AA Felder, M Schätz, B Schmerl, K Sugawara, C Prats, F Levet, A Klemm, K Miura, P Sampaio, C Tischer, R D'Antuono, BA Cimini, R Haase

\section*{Declaration of Interests}

K.S. is employed part-time by LPIXEL Inc.

\section*{Funding}

The following grants from the Chan Zuckerberg Initiative DAF, an advised fund of Silicon Valley Community Foundation, are reported: N.D.C. was supported as a CZI Imaging Scientist by grant number 2023-329684. A.F. was funded by grants 2022-309537 and 2024-349556, B.A.C. was supported by 2023-329649, C.G.W. by 2023-321238 to R.H.  B.A.C. was additionally supported by NIH P41 GM135019. L.R.B was supported by the Research Council of Norway through FRIPRO grant number 314684. B.S. is funded by DFG project 524025372 and additionally supported by NIH subaward RM1NS132981, and ONR grant N000142412055 to Elly Nedivi, and the Picower Institute Innovation Fund. This work was supported by the Francis Crick Institute, which receives its core funding from Cancer Research UK (CC1069), the UK Medical Research Council (CC1069) and the Wellcome Trust (CC1069). This work was supported by Vinicna Microscopy Core Facility (RRID:SCR$\_$026602) co-financed by the Czech-BioImaging large RI project LM2023050. The SciLifeLab BioImage Informatics Unit (A.C. and A.K.), a unit of the National Bioinformatics Infrastructure Sweden, receives funding from SciLifeLab and the Swedish Research Council (VR). P.S. was supported by grant FCT/EULAC/0002/2022.

\newpage
\subsection*{Supplementary}
\label{section:supplementary}

\renewcommand{\figurename}{Supplementary Figure}
\setcounter{figure}{0}

\begin{figure}[H]
    \centering
    \includegraphics[width=6.5in]{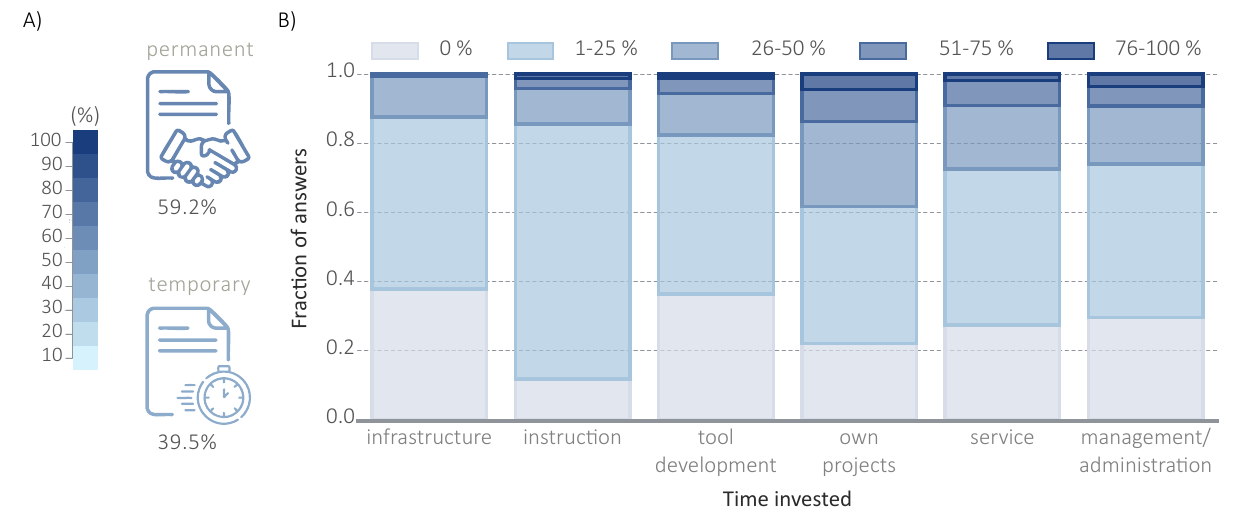}
    \caption{\textbf{Description of respondents current work. A)} Percentage of respondents with a permanent or temporary position. \textbf{B)} Stacked bar plot showing the percentage of time invested in different tasks as described by respondents.}
    \label{supplementary_figure:demographics}
\end{figure}

\newpage
\bibliographystyle{apalike}
\bibliography{./library}

\end{document}